\documentclass[aps,pra,twocolumn,showpacs]{revtex4}
\usepackage[pdftex]{graphicx}
\usepackage{amssymb}
\usepackage{amsmath}
\usepackage{bm}

\DeclareGraphicsExtensions{.pdf}

\begin{document}

\title{Periodic shedding of vortex dipoles from a moving penetrable obstacle in a Bose-Einstein condensate}
\author{Woo Jin Kwon, Sang Won Seo, Yong-il Shin}

\email{yishin@snu.ac.kr}

\affiliation{Department of Physics and Astronomy and Institute of Applied Physics, Seoul National University, Seoul 151-747, Korea}

\begin{abstract}
We investigate vortex shedding from a moving penetrable obstacle in a highly oblate Bose-Einstein condensate. The penetrable obstacle is formed by a repulsive Gaussian laser beam that has the potential barrier height lower than the chemical potential of the condensate. The moving obstacle periodically generates vortex dipoles and the vortex shedding frequency $f_v$ linearly increases with the obstacle velocity $v$ as $f_v=a(v-v_c)$, where $v_c$ is a critical velocity. Based on periodic shedding behavior, we demonstrate deterministic generation of a single vortex dipole by applying a short linear sweep of a laser beam. This method will allow further controlled vortex experiments such as dipole-dipole collisions. 
\end{abstract}


\maketitle

\section{Introduction}

Quantized vortices are the characteristic excitations of a superfluid and they play an important role in superfluid dynamics~\cite{Donnelly}. Since the first realization of atomic Bose-Einstein condensates (BECs), vortex dynamics has been one of the major focuses of BEC research~\cite{Fetter_rmp}. In recent experiments, many efforts were made to study the dynamics of vortex dipoles. A vortex dipole is a pair of vortices of opposite circulations and its creation and annihilation are at the heart of many two-dimensional (2D) superfluid phenomena such as Berezinskii-Kosterlitz-Thouless superfluids~\cite{B,KT}, phase transition dynamics~\cite{kibble,Zurek}, and superfluid turbulence~\cite{Takeuchi,White}. Quantized vortex dipoles were created in BECs by moving optical obstacles~\cite{inouye,neely,kwon1} and via quenching dynamics~\cite{anderson,hall}, and also observed in quasi-2D degenerate Bose gases~\cite{Choi_vp}. Orbital motions of a single vortex dipole in a trapped BEC were investigated~\cite{neely,hall,Middelkamp} and annihilation of vortex dipoles was indirectly probed in relaxation of superfluid turbulence in highly oblate BECs~\cite{kwon2}.

In this paper, we present an experimental study of vortex shedding from a moving repulsive Gaussian laser beam in a highly oblate BEC. The primary result of this work is that when the optical obstacle is penetrable (i.e., the potential barrier height $V_0$ of the laser beam is smaller than the chemical potential $\mu$ of the condensate), vortex dipoles are periodically shed from the moving obstacle. Furthermore, we find that the vortex shedding frequency linearly increases with the obstacle velocity as it exceeds a critical velocity. The periodic vortex shedding behavior is so pronounced that we demonstrate that a single vortex dipole can be deterministically generated by applying a short linear sweep of a penetrable laser beam. 

A fluid flowing past an obstacle is a textbook situation in fluid dynamics. In the case of a superfluid, it is well known that when the flow velocity exceeds a critical velocity, the superfluid becomes dissipative via vortex generation~\cite{frisch,jackson1}. Early theoretical works for 2D geometry anticipated periodic shedding of vortex dipoles from a moving circular obstacle~\cite{frisch,winiecki1,jackson1,winiecki2,jackson3}, where the periodic shedding was depicted as a cycling process where the work done by a drag force from the moving obstacle accumulates to generate a quantized vortex dipole of a finite energy. However, in previous BEC experiments~\cite{inouye,neely,kwon1}, periodic vortex shedding has never been observed, although the existence of a critical velocity was clearly demonstrated. 

In generating a vortex dipole, a penetrable obstacle with $V_0/\mu<1$ has a clear advantage over an impenetrable obstacle. Because there is no density-depleted region in the condensate, it is ensured that a vortex and an antivortex would be created at the same time into the condensate. On the other hand, in the impenetrable case, there is a density-depleted region occupied by the obstacle such that a vortex can be individually emitted into the condensate, while the other vortex that is created as a partner of the emitted vortex still remains in the density-depleted region~\cite{sasaki,saito2}. The individual vortex emission might be induced by small fluctuations in superfluid flow, braking the mirror symmetry of the system with respect to the obstacle moving direction. Recent numerical studies on vortex shedding from an impenetrable obstacle showed that vortex shedding is mostly irregular and regular shedding of vortex dipoles occurs only in a stringent condition of the obstacle~\cite{sasaki,reeves}, which is practically difficult to satisfy in current experiments. In this work, we observe that a moving \textit{penetrable} obstacle sheds vortex dipoles in a periodic and stable manner.

This paper is organized as follows. Sec.~II describes our setup and procedure of the vortex shedding experiment. Sec.~III presents the measurement results for various moving conditions of the obstacle and discusses the difference between the penetrable and impenetrable cases. Sec.~IV demonstrates an experimental method for deterministically generating a single vortex dipole and examines its stability in terms of of the linear velocity of the generated vortex dipole. Finally, Sec.~V provides a summary and outlook.

\begin{figure} 
\includegraphics[width=7.6cm]{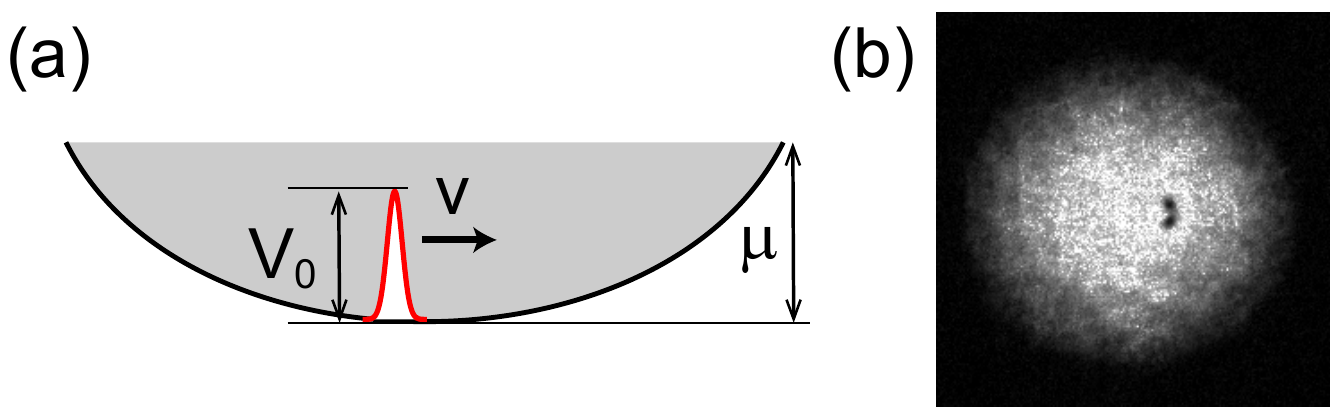}
\caption{(Color online) (a) Schematic of the experiment. A penetrable obstacle is formed by a repulsive Gaussian laser beam and it moves with velocity $v$ in the center region of a highly oblate Bose-Einstein condensate (BEC). The barrier height $V_0$ of the optical potential is lower than the chemical potential $\mu$ of the BEC. (b) Time-of-flight absorption image of a BEC. A vortex dipole is generated by the moving obstacle and it is identified with its pair of density-depleted cores.}
\label{setup}
\end{figure}

\section{Experiment}

The experimental setup is similar to that in our previous work~\cite{kwon1}. A highly oblate BEC of $^{23}$Na atoms is prepared in a hybrid trap composed of a pancake-shaped optical trap and a magnetic quadruple trap. The trapping frequencies are $\omega_{r,z}$ = $2 \pi \times (9.0, 400)$ Hz and the atom number in the condensate is $N_{0} = 3.3(2) \times 10^{6}$, giving the chemical potential $\mu \approx h \times 1.0$~kHz, where $h$ is the Planck constant.  The condensate fraction of the sample is over 80$\%$. At the trap center, the healing length is $\xi=\hbar/\sqrt{2m\mu} \approx 0.45~\mu$m and the speed of sound is $c_s=\sqrt{\mu/m}\approx 4.3~$mm/s, where $\hbar=h/2\pi$ and $m$ is the atomic mass. The Thomas-Fermi radius of the trapped condensate is $R=\sqrt{2\mu/m\omega_r^2} \approx 110~\mu$m. 

An optical obstacle is formed by a repulsive Gaussian laser beam whose $1/e^2$ width is  $\sigma=9.1(12)~\mu m \approx 20\xi$. We initially locate the laser beam 11.5~$\mu$m away from the condensate center and horizontally translate it through the center region of the condensate using a piezo-driven mirror~[Fig.~1(a)]. The moving velocity $v$ of the laser beam is kept constant during the translation. After the laser beam sweep, we linearly ramp down the beam power within 20~ms and then take an absorption image of the condensate to detect vortices~[Fig.~1(b)]. When we release the trapping potential, we turn off the magnetic trap 13~ms earlier than the optical trap and place 24~ms free expansion before taking the image.

\begin{figure} 
\includegraphics[width=7.8cm]{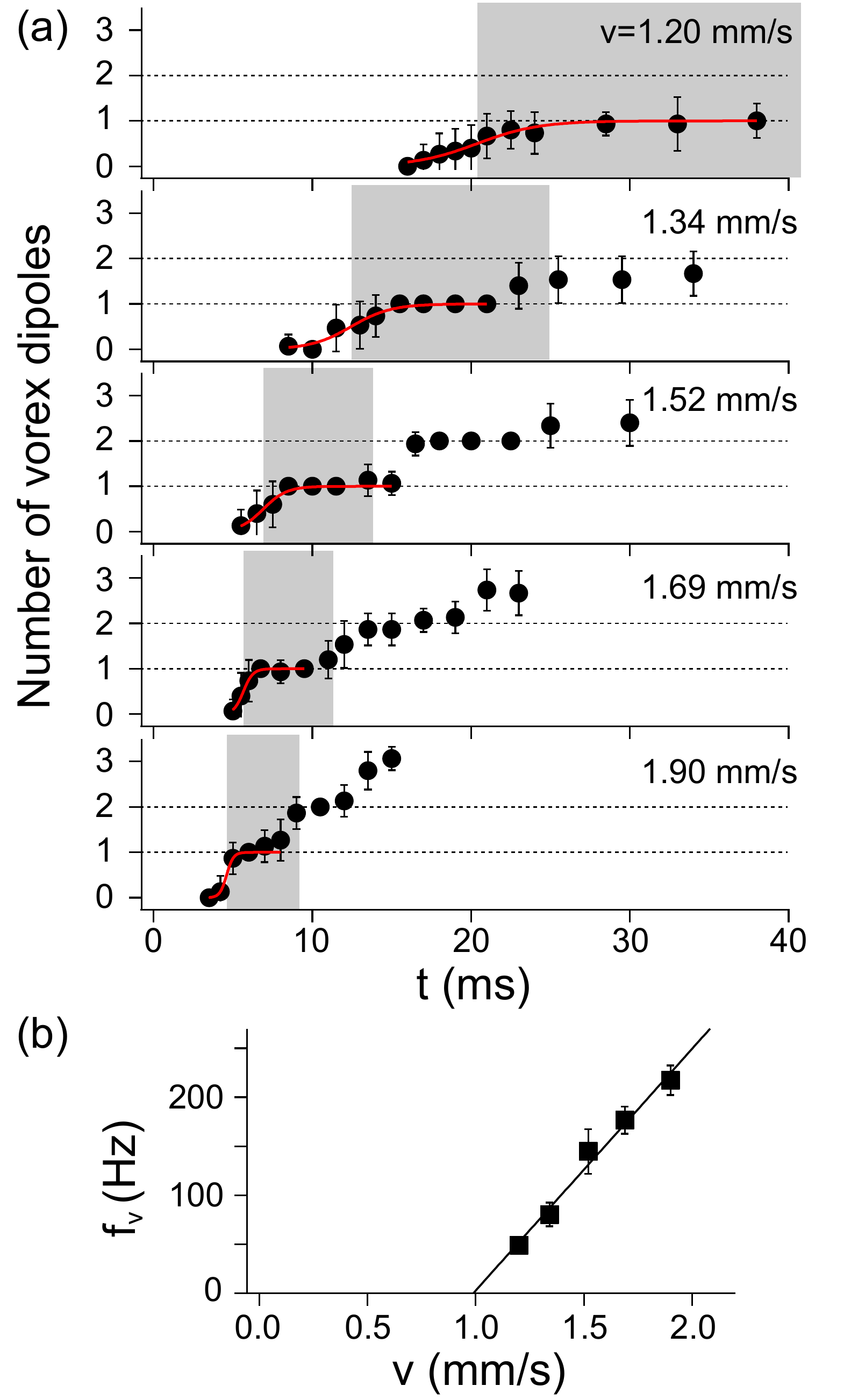}
\caption{(Color online) Periodic shedding of vortex dipoles from a moving penetrable obstacle of $V_0/\mu\approx 0.74$. (a) Number of generated vortex dipoles, $N_d$ as a function of the sweeping time $t$ for various moving velocities $v$ of the obstacle. Each data point consists of 15 realizations of the same experiment and the error bar indicates the standard deviation of the measurements. The activation time $\tau_v$ for generating the first vortex dipole was determined from a sigmoidal function fit (red solid lines, see the text for detail) to the data up to the $N_d$=1 plateau region. The shaded area indicates the region of $\tau_v<t<2\tau_v$. (b) Vortex shedding frequency $f_v=1/\tau_v$ as a function of $v$. The solid line is a linear fit to the data.}
\label{Fig2}
\end{figure}

\section{Periodic vortex shedding}

\subsection{Penetrable obstacle}

We first investigate the case of a penetrable obstacle with $V_0/\mu \approx 0.74$ by measuring the number of vortex dipoles generated from the obstacle as a function of the sweeping time $t$ for various moving velocities $v$. The measurements results are displayed in Fig.~2(a). We see that the vortex dipole number $N_d$ shows stepwise increasing behavior as a function of $t$. There is a certain amount of an inert time $\tau_v$ before generating the first vortex dipole and another sweeping time almost equal to $\tau_v$ is required to generate the second vortex dipole. This regular stepping-up feature clearly indicates that the inert time is not due to the transient response of the condensate to the perturbations of the laser beam but an intrinsic activation time for generating a vortex dipole from the moving obstacle. The periodic vortex generation might be understood as a cycling process where the moving obstacle exerts a drag force during the activation time and the accumulated energy is released as a vortex dipole at a certain threshold condition~\cite{winiecki1}.

\begin{figure} 
\includegraphics[width=7.4cm]{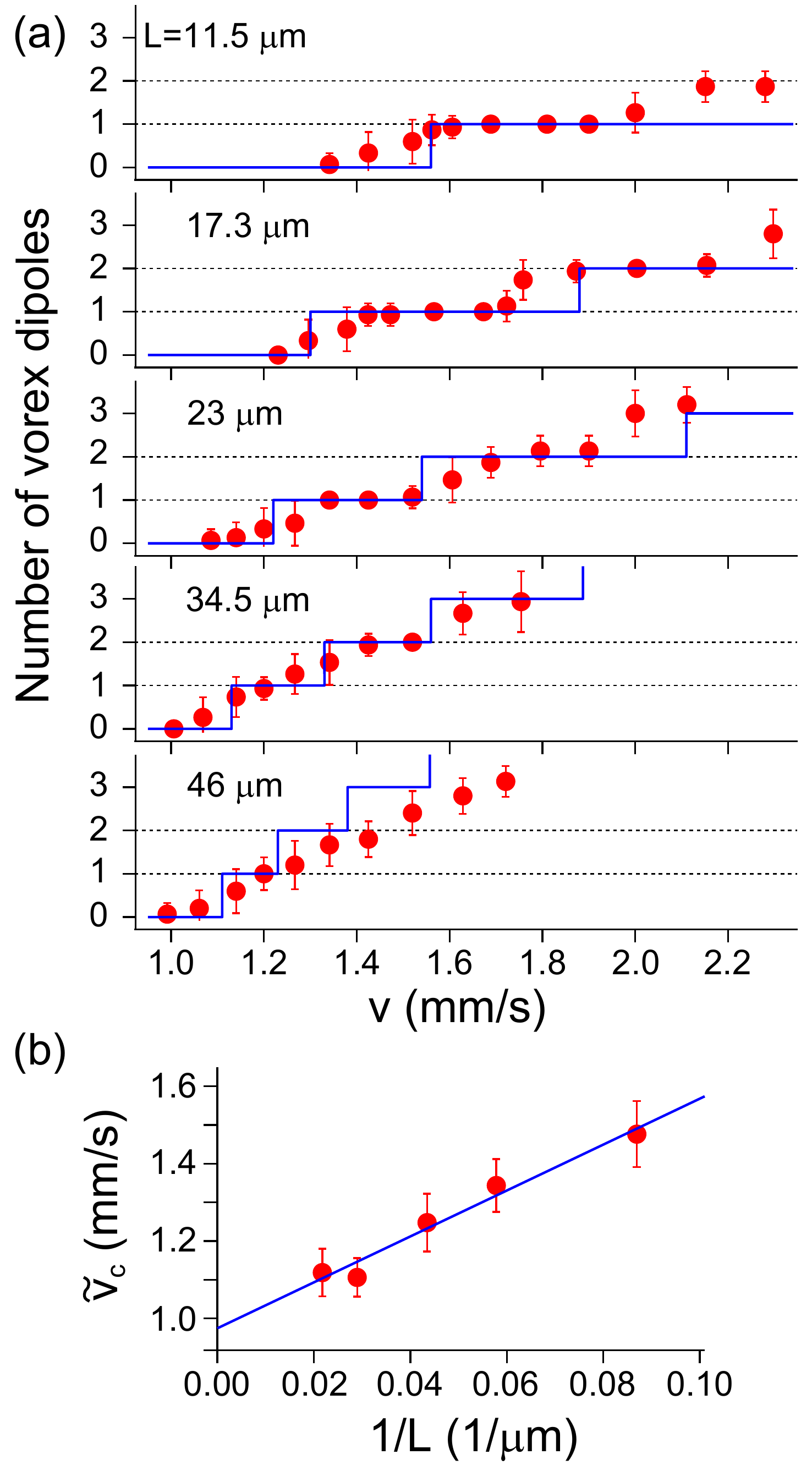}
\caption{(Color online) (a) Vortex dipole number $N_d$ as a function of $v$ for fixed sweep length $L=vt$. $V_0/\mu\approx 0.74$. The blue solid lines are the theoretical predictions obtained from a model assuming perfect periodic vortex shedding with $f_v=a(v-v_c)$, $a=0.25~\mu$m$^{-1}$, and $v_c=1.0$~mm/s [Fig.~2(b)]. (b) Critical velocity $\tilde{v}_c$ measured for fixed $L$. $\tilde{v}_c$ was determined from the data in (a), using the data analysis method described in Ref.~\cite{kwon1}. The solid line is a linear fit to the data.}
\label{Fig3}
\end{figure}

It was theoretically anticipated that the vortex shedding frequency $f_v$ linearly increases with the moving velocity $v$ of the obstacle as $f_v=a(v-v_c)$, where $v_c$ is a critical velocity~\cite{frisch,winiecki1,winiecki2,jackson3,tsubota2,footnote1}. Interpreting the activation time $\tau_v$ as the inverse of $f_v$, we find that our experimental data support the theoretical prediction~[Fig.~2(b)]. $\tau_v$ was determined by fitting a sigmoidal function $N_d=[1+e^{-(t-\tau_v)/\gamma}]^{-1}$ to the data up to the $N_d$=1 plateau region. Here $\gamma$ represents the jittering of the vortex emission event and we used the value of $1.5\gamma$ as the measurement uncertainty of $\tau_v$. Fig.~2(b) shows $f_v=1/\tau_v$ as a function of $v$ and the result is well described with the functional form of $f_v=a(v-v_c)$, giving $v_c=0.99(4)$~mm/s and $a=0.25(2)~\mu$m$^{-1}$.

We present another set of measurements in Fig.~3(a), where $N_d$ is measured as a function of $v$ for fixed sweep lengths $L=vt$ of the laser beam. This is a typical situation addressed in previous experiments to measure the critical velocity $v_c$ for vortex shedding~\cite{neely,kwon1,kwon2}. Our motivation is to examine the effect of the finite activation time $\tau_v$ on the determination of $v_c$. In order to generate vortices, the sweeping time $t=L/v$ should be longer than $\tau_v$, which requires $v<aL(v-v_c)$ from the relation $f_v=a(v-v_c)=1/\tau_v$. Therefore, in the fixed $L$ setting, the critical velocity would be measured with a systematic error of $\delta v_c=v_c/(aL-1)$. Indeed, we observe that the measured critical velocity $\tilde{v}_c$ increases with decreasing $L$ [Fig.~3(b)]. $\tilde{v}_c$ was determined using the data analysis method described in our previous work~\cite{kwon1}. The inhomogeneous density effect is negligible because the variation of the local speed of sound $c_s$ is less than 5\% even for the longest $L=46~\mu$m. The linear extrapolation of the experiment data to $1/L=0$ gives a velocity of 0.97(3)~mm/s [Fig.~3(b)], which is remarkable in that this value is consistent with the critical velocity $v_c$ obtained from the vortex shedding frequency data in Fig.~2(b).

The blue solid lines in Fig.~3(a) indicate the model predictions assuming perfect periodic generation with $f_v=a(v-v_c)$, $a=0.25~\mu$m$^{-1}$ and $v_c=1$~mm/s. Most of the qualitative features of the experimental data are explained with the model, including the reduction of the $N_d$=1 plateau region with increasing $L$. This observation corroborates the periodic shedding of vortex dipoles from the moving penetrable obstacle. 

\subsection{Impenetrable obstacle}

We perform the same measurements with an impenetrable laser beam of $V_0/\mu \approx 1.5$ (Fig.~4). Interestingly, the activation time is still necessary for generating the first vortex dipole and reduces with higher moving velocity as observed with the penetrable obstacle. This seems to be consistent with our description that the activation time is required to accumulate a certain amount of energy with the drag force to form a vortex dipole. However, no periodic vortex shedding is observed with the impenetrable obstacle and the number of vortex dipoles grows monotonically once the first vortex dipole is emitted. This means that the subsequent vortex generations are significantly affected by the presence of the first vortex dipole~\cite{saito2,berloff2}. It is noted that for $v=1.90$~mm/s in Fig.~4(a), the increasing rate of $N_d$ is almost four times higher than the inverse of the activation time, which implies that vortex shedding can be accelerated in the presence of other vortices near the obstacle.

\begin{figure} 
\includegraphics[width=7.3cm]{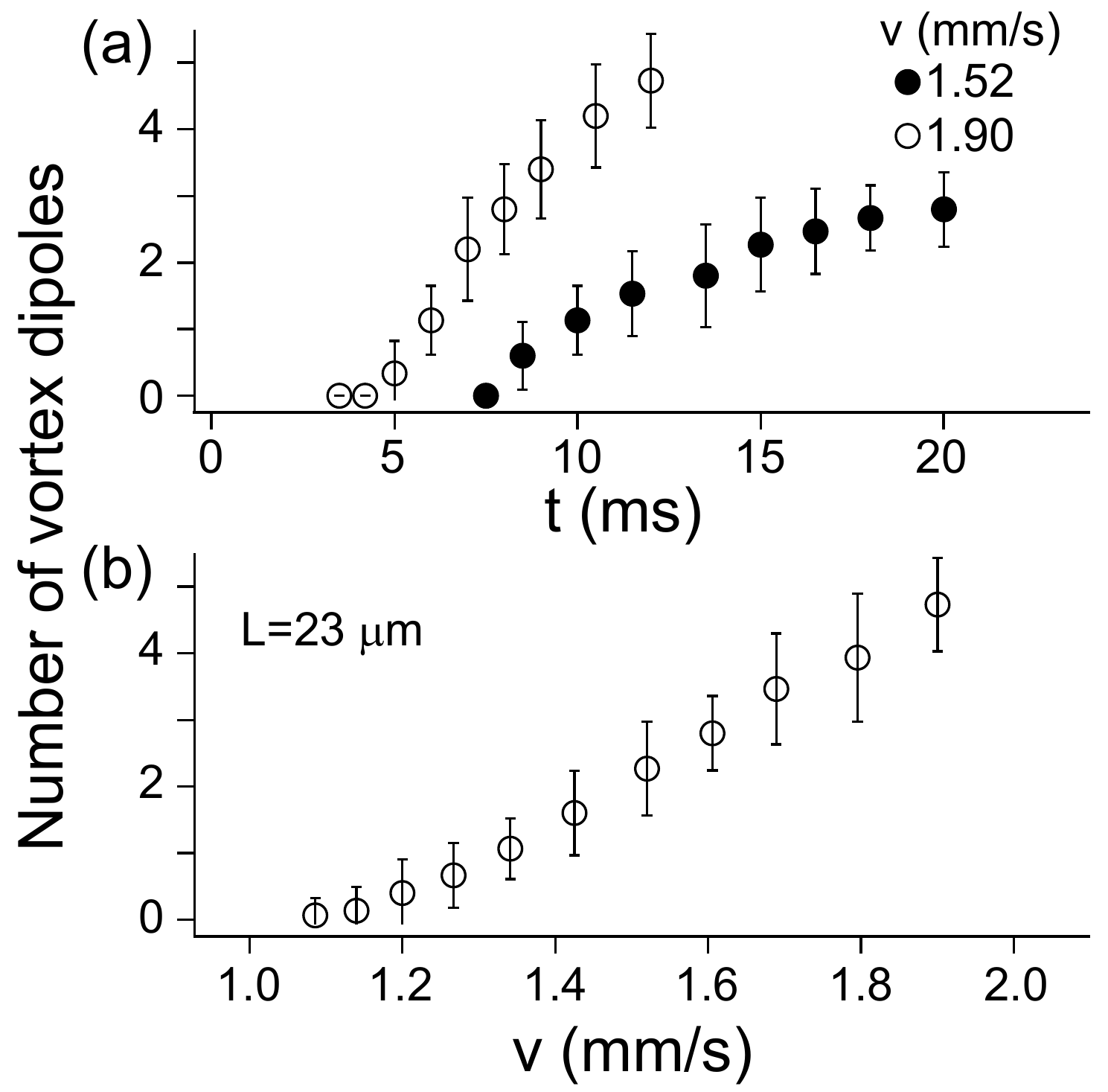}
\caption{Vortex shedding from a moving impenetrable obstacle of $V_0/\mu \approx 1.5$. (a) Vortex dipole number $N_d$ versus sweeping time $t$. (b) $N_d$ as a function of the moving velocity $v$ for sweep length $L=23~\mu$m.}
\label{Fig4}
\end{figure}

The distinctive difference of the impenetrable case from the penetrable case is that there is a density-depleted region occupied by the obstacle. Hence, it is possible for vortices to be emitted individually into the condensate. The individually emitted vortices would disturb the superfluid velocity field around the obstacle more strongly than a vortex dipole, leading to irregular vortex shedding or possibly more intriguing shedding patterns such as the B\'enard-von K\'arm\'an vortex street~\cite{sasaki,saito2}.

In Fig.~5, we present several vortex images containing a few vortex dipoles. It is clearly seen that the vortex shedding pattern is irregular in the impenetrable case and was often observed that the propagation direction of the first vortex dipole largely deviated from the moving direction of the obstacle [Fig.~5(a)], implying that the two vortices were emitted at different times from the obstacle. Furthermore, vortex clusters having larger density-depleted cores were sometimes observed [Fig.~5(c)]~\cite{neely,reeves}. On the other hand, the first vortex dipole from a penetrable obstacle usually propagates along the moving direction of the obstacle. Symmetric periodic array of vortex dipoles are known to be unstable towards staggered configurations of vortex dipoles~\cite{sasaki,saito2,nore}, which were typically observed in the experiment when multiple vortex dipoles were generated~[Figs.~5(d) and (e)]. Occasionally, symmetric trains of vortex dipoles were observed~[Fig.~5(f)].

\begin{figure} 
\includegraphics[width=6.8cm]{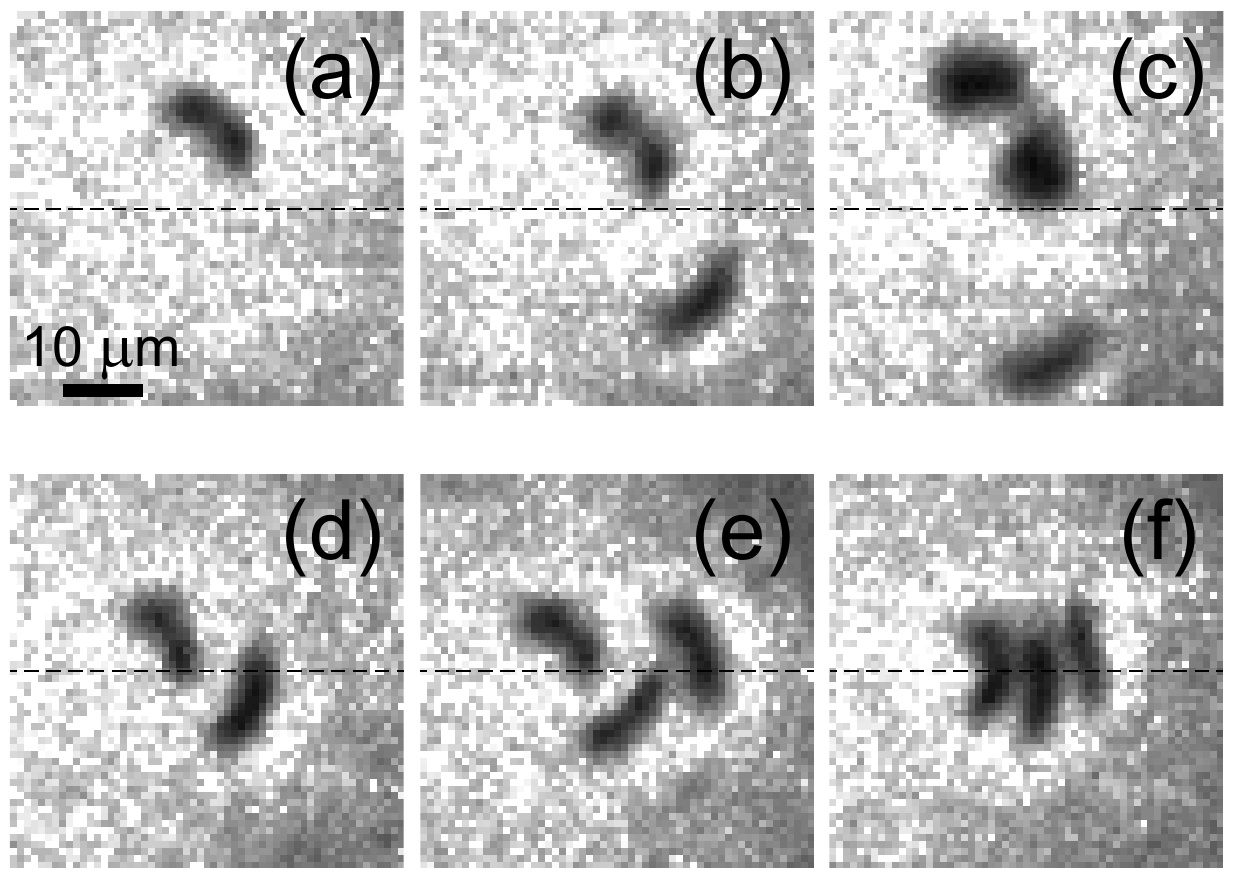}
\caption{Various patterns of vortex shedding from (a)-(c) impenetrable and (d)-(f) penetrable obstacles. In the impenetrable case, the first vortex dipole was often observed to propagate in a direction largely deviated from the +$x$ moving direction of the obstacle. The dashed lines indicate the moving line of the obstacle. (c) Vortex clusters having larger density-depleted regions  were also observed. In the penetrable case, staggered configurations of multiple vortex dipoles were typically observed. (f) Occasionally, symmetric trains of vortex dipoles also appeared.} 
\label{Fig7}
\end{figure}

\section{Deterministic generation of a single vortex dipole}

In the experiment data for the penetrable obstacle in Fig.~2(a), it is evident that fluctuations of the vortex dipole number $N_d$ are strongly suppressed for $\tau_v<t<2 \tau_v$ and even vanish for some conditions. This is an outstanding feature of the periodic vortex shedding, suggesting that deterministic generation of a single vortex dipole is possible when applying a short sweep of the penetrable laser beam. In this section, we characterize this short sweep method by examining the stability of the propagation velocity of the generated vortex dipole.

In Figs.~6(a) and (b), we show two exemplary images for $v=1.52$~mm/s with $t=11.5$~ms and 7.5~ms, respectively. In the first image, a vortex dipole is located near the center region and clearly identified with its two distinctive density-depleted cores. On the other hand, in the second image that is taken with a short sweeping time $t\sim\tau_v$, a small crescent-shaped density dimple is observed with low visibility in the boundary region of the condensate. We interpret this density dimple as an incipient vortex dipole with a short intervortex distance $d$, which cannot be fully developed because of the short sweeping time. The incipient vortex dipole moves further away from the sweeping region because of higher propagation speed $v_d=\hbar/md$ for smaller $d$ and its low visibility might be attributed to the blurring effect during the expansion before imaging.

\begin{figure}
\includegraphics[width=7cm]{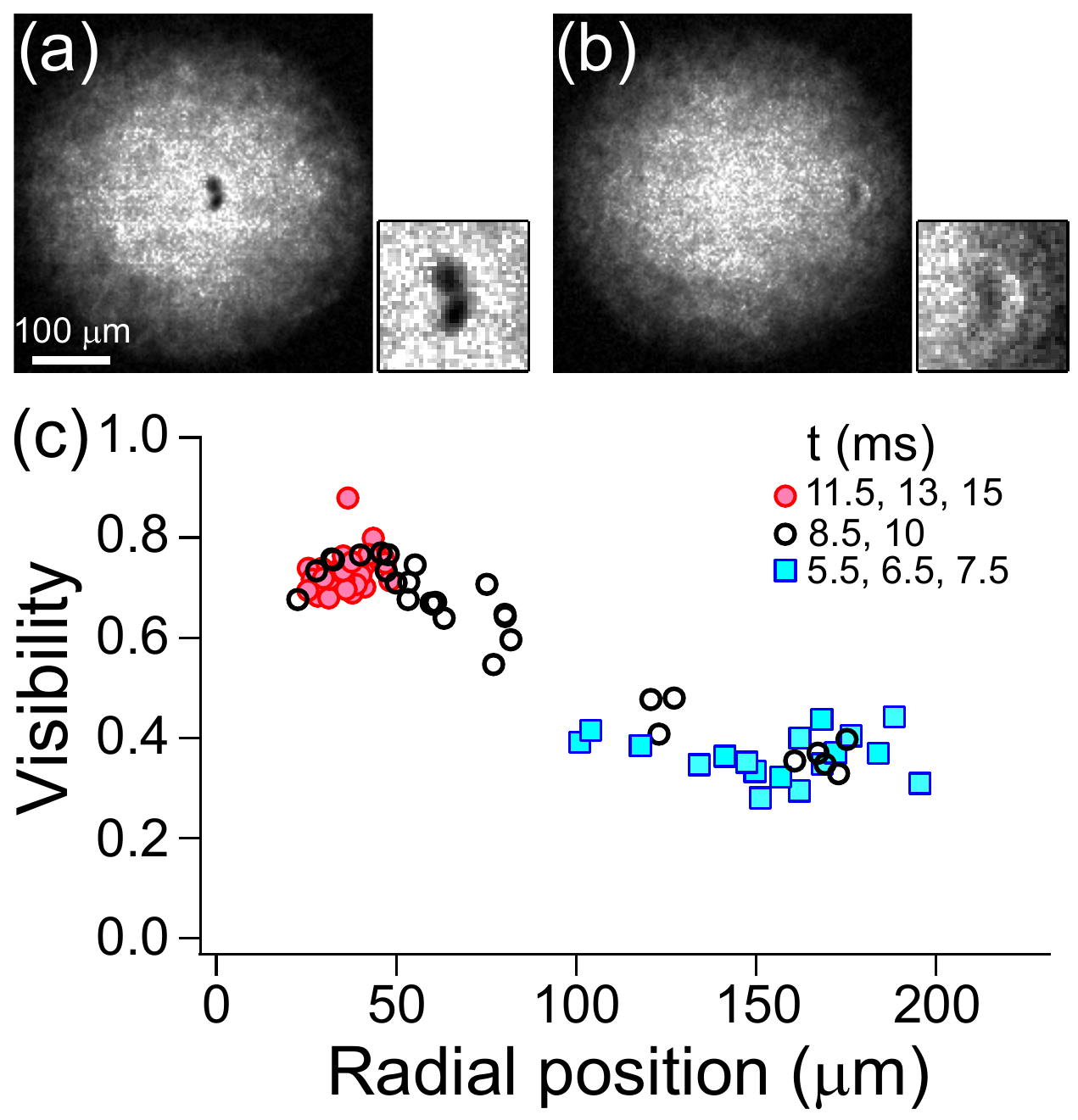}
\caption{(Color online) Formation of a vortex dipole. Images of BECs containing a vortex dipole, where $V_0/\mu\approx 0.74$ and the sweeping times were (a) $t=11.5$~ms and (b) 7.5~ms at $v=1.52$~mm/s, respectively. For long $t$, the vortex dipole is clearly identified with its two vortex cores near the center region. For short $t$, it appears as a small density dimple with low visibility in the boundary region. The enlarged images of the vortex dipoles are presented on the right side of each image. (c) Vortex distribution in the plane of radial position and visibility, obtained from all the images containing a single vortex dipole in the experiment for $v=1.52$~mm/s [Fig.~2(a)].}
\label{Fig6}
\end{figure}

Compiling all the images containing a single vortex dipole for $v=1.52$~mm/s, in Fig.~6(c) we display the distribution of the vortex dipoles in the plane of radial position and visibility. The visibility was determined as $1-n_m/\bar{n}$, where $n_m$ is the density minimum value in the core region and $\bar{n}$ is the average density in the surrounding area of the vortex dipole. The distribution shows a strong correlation between the visibility of the vortex dipole and its radial position, i.e., propagation speed, revealing the temporal development of the vortex dipole. From a rough estimation, taking into account the expansion time in the imaging sequence, we find that the propagation speed of the most incipient vortex dipole is comparable to the moving velocity of the obstacle, which is expected from the fact that the vortex dipole is shed from the moving obstacle. The condition of $v_d=v$ gives the intervortex distance $d=\hbar/mv\approx 4\xi$.

When the sweeping time is not sufficiently long, although long enough to ensure generation of a vortex dipole, the speed of the generated vortex dipole is not reproducible. For example, when $t=8.5$ or 10~ms, the vortex generation probability is almost unity but the vortex dipole velocity widely scatters [open circles in Fig.~6(c)]. This is partially due to the uncertainty of the vortex emission time and also probably due to the finite dwelling time of the vortex dipole in the proximity of the laser beam, which might be estimated to be $\Delta t=\sigma/2v$. Therefore, for deterministic generation of a single vortex dipole, the sweeping time needs to be optimally chosen to have better stability in the vortex dipole velocity $v_d$. We examined the stability of $v_d$ by putting an additional 100~ms hold time before releasing the trapping potential and measuring the change in the vortex position. In our experiment, the propagation speed of the vortex dipole was stable within 10\% and its propagation direction was parallel within 15 degrees to the sweeping direction of the laser beam.

It needs to be mentioned that we cannot exclude the possibility that the crescent-shaped density dimple is a rarefaction pulse or a gray soliton, which can also propagate with preserving its density profile with a nonzero density minimum~\cite{tsubota2,berloff1,saito1,huang}. When the energy accumulated by the moving obstacle is not sufficient to generate a vortex dipole, it can possibly transform into lower energy excitation. The sudden stop of the moving obstacle might play a role in forming and stabilizing the excitation.

\begin{figure}
\includegraphics[width=8.4cm]{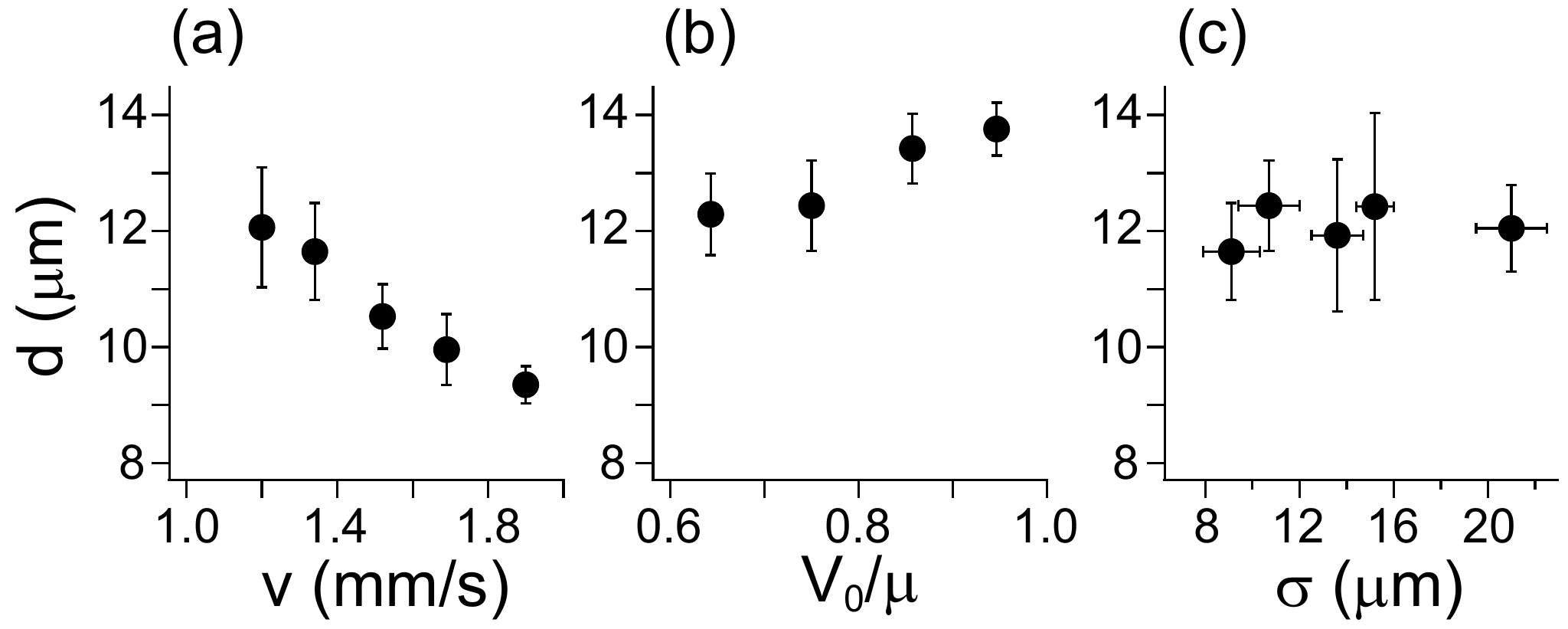}
\caption{Dependence of the intervortex distance $d$ on various sweeping parameters of the penetrable obstacle. (a) $d$ versus $v$, where $\sigma=9.1(12)~\mu$m and $V_0/\mu\approx 0.74$. (b) $d$ versus $V_0/\mu$, where $\sigma=10.7(13)~\mu$m and $v=1.25$~mm/s. (d) $d$ versus $\sigma$, where $V_0/\mu\approx 0.74$. In (d), $v$ was controlled to range from 1.08~mm/s to 1.34~mm/s due to the change of the critical velocity~\cite{kwon1}. In each measurement, the sweeping time $t$ was chosen to have $d$ stable (see the text for detail). $d$ was measured in the image and divided by the scaling factor of 2.1 for the expansion in the imaging sequence.}
\label{Fig5}
\end{figure}

Finally, with a motivation to find a way to control the velocity of the vortex dipole, we investigate the dependence of the intervortex distance $d$ on various parameters of the moving obstacle. Fig.~7 displays the measurement results of $d$ as functions of the moving velocity $v$, the barrier height $V_0/\mu$, and the beam waist $\sigma$ of the laser beam. We see that $d$ slightly decreases with increasing $v$, suggesting one way to control $v_d$ but in a limited range. Intriguingly, $d$ is almost insensitive to $V_0/\mu$ and $\sigma$. Considering the fact that the critical velocity $v_c$ is significantly affected by the potential parameters of the obstacle~\cite{kwon1}, the observed weak dependence of $d$ on them is quite unexpected. In Ref.~\cite{saito1}, vortex shedding from a moving attractive obstacle of $V_0/\mu<0$ was numerically studied, showing that $d$ increases with higher $v$, which is opposite to our observation. Further theoretical studies are warranted to understand what determines the size of the vortex dipole when it is generated from a moving penetrable obstacle.

\section{Summary and Outlook} 

We investigated vortex shedding dynamics with a moving penetrable obstacle in a highly oblate BEC. We observed periodic shedding of vortex dipoles and found the linear dependence of the vortex shedding frequency on the moving velocity of the obstacle above the critical velocity. In addition, we demonstrated deterministic generation of a single vortex dipole by applying a short linear sweep of the penetrable laser beam. This vortex generation method is expected to provide new opportunities for further controlled experiments on vortex dynamics. For example, as suggested in Ref.~\cite{saito1},  if two laser beams are employed to generate two vortex dipoles separately at different positions in a condensate, it would be possible to investigate collision dynamics of vortex dipoles in a controlled manner. Dipole-dipole collisions are particularly interesting in that vortex pair annihilation may occur  during the collision. Vortex pair annihilation is one of the key issues in the study of 2D quantum turbulence~\cite{onorato,horng,tsubota1,schole,chesler1,simula,stagg,chesler2,zhang}. Many aspects of the pair annihilation process~\cite{rorai,angom}, such as the roles of vortex-phonon interactions~\cite{dutton,lucas}, and thermal dissipation~\cite{allen}, have never been directly addressed in experiments.

\begin{acknowledgements} 

We thank Seji~Kang for experimental assistance. This work was supported by the National Research Foundation of Korea (Grant No. 2011-0017527).

\end{acknowledgements}


\begin{references}
\bibitem{footnote1} There is a theoretically subtle issue on whether the vortex nucleation period is finite~\cite{winiecki1} or diverges~\cite{tsubota2} as $v$ converges to $v_c$.
\end{references}
\end{document}